\documentclass[preprint,aps,prl,showpacs,showkeys,superscriptaddress]{revtex4-1}
\usepackage[english]{babel}
\usepackage{tabulary,graphicx,amsmath,amsfonts,amssymb}
\usepackage[utf8x]{inputenc}
\usepackage{tikz}
\usepackage{braket}
\usepackage{pgfplots}
\usepackage{csquotes}
\usepackage{hhline}

\usepackage{dcolumn}
\usepackage{tabularx}
\usepackage[colorlinks=true,linkcolor=blue,citecolor=blue,urlcolor=blue]{hyperref}
\usepackage{longtable}
\usepackage{float}
\usepackage{epsfig}

\usepackage{url,multirow,morefloats,floatflt,cancel,tfrupee}
\makeatletter

\AtBeginDocument{\@ifpackageloaded{textcomp}{}{\usepackage{textcomp}}}
\makeatother
\usepackage{colortbl}
\usepackage{xcolor}
\usepackage{pifont}
\usepackage[nointegrals]{wasysym}
\makeatletter

\newcolumntype{C}{>{\centering\arraybackslash}X}
  \pgfplotsset{compat=1.14}
  
\begin{document}
\setcounter{secnumdepth}{3}
\title{Quantum Locker Using a Novel Verification Algorithm and Its Experimental Realization in IBM Quantum Computer}

\author{Avinash Dash}
\email{ad16ms036@iiserkol.ac.in} 
\affiliation{Department of Physical Sciences, Indian Institute of Science Education and Research Kolkata, Mohanpur, 741246, West Bengal, India}

\author{Sumit Rout}
\email{rsumitrout3@gmail.com}
\affiliation{Integrated Science Education and Research Centre, Visva Bharati University, Santiniketan, 731235, West Bengal, India}

\author{Bikash K. Behera}\email{bkb13ms061@iiserkol.ac.in}
\author{Prasanta K. Panigrahi}
\email{pprasanta@iiserkol.ac.in}
\affiliation{Department of Physical Sciences, Indian Institute of Science Education and Research Kolkata, Mohanpur, 741246, West Bengal, India}

\begin{abstract}

\textbf  {
          It is well known that Grover's algorithm asymptotically transforms an equal superposition state into an eigenstate (of a given basis). Here, we demonstrate a verification algorithm based on weak measurement which can achieve the same purpose even if the qubit is not in an equal superposition state. The proposed algorithm highlights the distinguishability between any arbitrary single qubit superposition state and an eigenstate. We apply this algorithm to propose the scheme of a Quantum Locker, a protocol in which any legitimate party can verify his/her authenticity by using a newly developed quantum One-Time Password (OTP) and retrieve the necessary message from the locker. We formally explicate the working of quantum locker in association with the quantum OTP, which theoretically offers a much higher security against any adversary, as compared to any classical security device.
          }

\end{abstract}


\maketitle

\section{Introduction}

Grover's search algorithm is the optimal algorithm for achieving quadratic speedup in searching a particular ``marked" state from an uniformly distributed database \cite{219067:4918030}.  A generalization of the algorithm \cite{219067:4918149, 219067:4918839, 219067:4918978, 219067:4919007, 219067:4918245, 219067:4919049, 219067:4918881, 219067:4919155, 219067:4919157, 219067:4919441} allows for any arbitrary superposition state of the quantum database instead of the equal superposition state \cite{219067:4918149, 219067:4918245}. Using the standard Grover's algorithm, a marked state can not be obtained from the initial state of the quantum database with $100\%$ success. However, the same can be achieved with certainty by a modification of Grover's algorithm as proposed by Hoyer \cite{219067:4918839}. Long also presents a modified search algorithm with a zero theoretical failure rate \cite{219067:4918881}, where the algorithm searches a marked state from an evenly distributed database using phase matching condition \cite{219067:4919007, 219067:4918978}. Liu \cite{219067:4919049} generalized Long's algorithm, in which multimarked states are searched with certainty from an arbitrary $N$-item complex initial amplitude distributed quantum database ($N$ is not necessarily $2^n$).

In the scheme proposed by Liu, a system of multimarked states can be deterministically searched from an arbitrary initial state of an unsorted database, of any arbitrary size. Let us consider a particular case, in which we intend to extract the state $\Ket{1}$ from the database having arbitrary superposition state $\alpha\Ket{0}+\beta\Ket{1}$ ($\alpha,\beta\in\mathbb{C}$, $\beta\neq0$). In order to do so, the machine that executes the algorithm must have \textit{a priori} knowledge of the coefficients $\alpha$ and $\beta$. But what if these coefficients are unknown? Liu's scheme fails to overcome this issue, which has been tackled by our proposed algorithm which can search a particular marked state from an arbitrary superposition state with asymptotic determinism, even if the initial state is unknown to the machine.

In the above example, if we take the initial database state to be the state $\Ket{0}$ ($\beta=0$), then Liu's scheme fails altogether, since one of the algorithm's parameters becomes indeterminant. However, the database will remain in the state $\Ket{0}$ if our algorithm is operated. Hence, besides being a quantum search algorithm, this also serves as an algorithm for verification, as it can asymptotically distinguish a single qubit eigenstate of any basis from an unknown and arbitrary single qubit superposition state. We apply this novel result of the algorithm to propose a scheme of a quantum locker for which the motivation is described as follows.

Before the advent of electronic information storage and transfer, a classical locker with a physical key was the only means available for secure storage of contents. The need to transmit information securely called for advanced methods of user authentication, \textit{e.g.} passwords \cite{219067:4929851}. The onset of the digital era has consequentially paved the way for highly secure means of information storage and retrieval such as e-lockers. However, the digitization era also brought with it increased vulnerability of static passwords to replay attacks \cite{219067:4929852}. Recently, one-time passwords (OTPs) \cite{219067:4929853} based on pseudo-random numbers have gained popularity, which are time-limited and suitable for highly confidential transactions.

Over the past few years, the focus has shifted towards the development of security protocols that exploit some of the exclusive features of quantum mechanics, which may offer higher security than their classical counterparts.~One such feature is the `no-cloning theorem' \cite{219067:4919523}, which forbids the construction of an exact replica of a generic/unknown quantum state. This property is greatly beneficial in quantum cryptosystems as it enables two communicating parties to detect whether or not an adversary has intercepted the transmitted message. Many quantum security schemes have been devised which have proved to be unconditionally secure protocols for safeguarding information \cite{219067:4919565, 219067:4919569, 219067:4919611, 219067:4919614, 219067:4919615, 219067:4921045}. Such protocols effectively use quantum phenomena like superposition and entanglement \cite{219067:4929854}. Some of these information security techniques include a quantum key distribution scheme \cite{219067:4919565, 219067:4919569, 219067:4919611, 219067:4919614, 219067:4919615, 219067:4921045, 219067:4921046, 219067:4921047}, quantum identification scheme \cite{219067:4921048, 219067:4921049, 219067:4921050}, quantum digital signature scheme \cite{219067:4921051}, quantum cheque scheme \cite{219067:4921052, 219067:4921053}, to name a few.

There have been many works in the past few years regarding the realization of OTPs in quantum format.~A quantum one-time password based on a set of $n$ entanglements (Bell States) was proposed by Mihara \cite{219067:4921054}. Besides, recent works have led to the realization of robust authentication protocols based on quantum passwords \cite{219067:4921055, 219067:4921056}. Recently, encrypted data storage by the application of a disordered field on photonic quantum memories has been accomplished \cite{219067:4921057}.

The proposed quantum locker is one of many applications realizable in the future which will practically use quantum passwords. In our proposed scheme, the locker does not involve any classical operations, nor does it require bits to store the output. In the course of our work, we hold the assumption that the sharing of entanglement between the two communicating parties considered is secure and can be preserved indefinitely without getting decohered or otherwise lost by other means \cite{219067:4921058, 219067:4919565}. All quantum channels and gates used in our scheme are assumed to be free from decoherence and errors.

Recently, a series of quantum information processing tasks \cite{219067:4920990, 219067:4921059, 219067:4920991, 219067:4920992, 219067:4920993, 219067:4920994, 219067:4920995, 219067:4920996, 219067:4920997, 219067:4920998, 219067:4920999, 219067:4921000, 219067:4921001, 219067:4921002, 219067:4921003} have been run using IBM quantum computer. Hence, motivated by this fact, we have used IBM's 5 qubit quantum processor `ibmqx4' to carry out the experimental procedure to explicitly show the working of our proposed verification algorithm and demonstrate the scheme of a Quantum Locker.       

\section{Results}

\textbf{Verification Algorithm \label{qlock_veri}}

\begin{figure}[!ht]
\centering
\includegraphics[width=\linewidth]{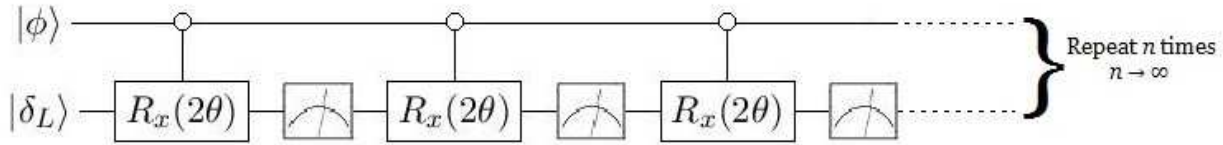}
\caption{\emph{Circuit illustrating the Verification Box $V$.}}
\label{qlock_Figure1}
\end{figure}
 
This algorithm uses a verification box $V$ (Fig. \ref{qlock_Figure1}) to discriminate two non-orthogonal states, i.e., $\ket0$, and a generic state $\ket\phi\equiv\alpha\ket0+\beta\ket1$. The proposed scheme is composed of several iterations of a weak measurement protocol \cite{219067:4929855}, where an ancillary qubit $\ket{\delta_L}$ (in the state $\ket0$) is weakly coupled to the system. After each iteration, a projective measurement is performed on the ancilla, which slightly perturbs the state of the system ($\ket\phi$). The evolution of the coupled system ($\ket\phi\otimes\ket{\delta_L}$) after each iteration is described by the following unitary operator $U$,
\begin{equation}
\label{qlock_1}
U=[R_z(\theta)\otimes I_{2}][cos\theta I_{4}-isin\theta C^0NOT_{12}]    
\end{equation}
where $I_{2}$ and $I_{4}$ denote identity matrices of order 2 and 4 respectively. Here, $C^0NOT_{12}$ flips the target qubit only when the control qubit is in state $\ket0$. It is to be noted that the operator $U$ describing the evolution of the coupled system is simply the $Controlled^0-R_x(2\theta)$ operation, where the operator $R_x(2\theta)$ acts on the ancilla (target) qubit only if the system (control) qubit is in state $\ket0$. Here, $\theta$ is a parameter intrinsic to the locker, which is an arbitrarily small value ($\theta\rightarrow$0). At the end of the first iteration, the operator $U$ transforms the composite system $\ket\phi\ket{\delta_L}$ into the following state, 

\begin{equation}
U\ket\phi|\delta_L\rangle\equiv (\alpha cos\theta\ket0+\beta\ket1)\ket0-\alpha isin\theta\ket0\ket1
\end{equation}

\begin{figure}[!ht]
\centering
\includegraphics[width=\linewidth]{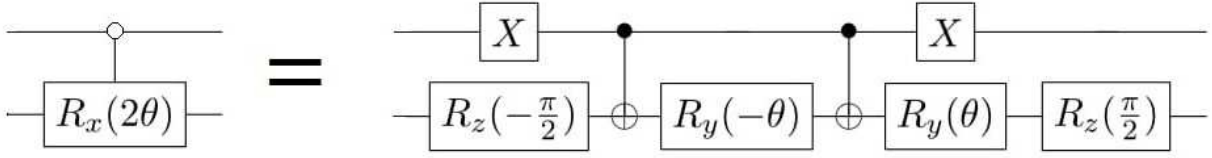}
\caption{\emph{Circuit depicting the implementation of $U\equiv Controlled^0-R_x(2\theta)$.}}
\label{qlock_Figure2}
\end{figure}

It can be verified that the operation $Controlled^0-R_x(2\theta)$ can be implemented by using universal single qubit gates and $CNOT$ gate, as shown by the following circuit (Fig.~\ref{qlock_Figure2}).

After the execution of the unitary transformation $U$, a projective measurement (in \textit{z}-basis) is performed on the ancillary qubit, resulting in outcomes $\ket0$ and $\ket1$ with probabilities $p_0$ and $p_1$ respectively.
\vspace{-.35cm}
\begin{eqnarray}
&& \label{qlock_5}p_0=|\alpha|^2cos^2\theta+|\beta|^2\approx1-|\alpha|^2\theta^2\\
&& \label{qlock_6}p_1=|\alpha|^2sin^2\theta\approx|\alpha|^2\theta^2
\end{eqnarray}

In the limit $\theta\rightarrow0$, the probability of obtaining the state $\ket1$ on measurement of the ancilla is negligible. Hence, assuming the outcome of the ancilla state to be $\ket0$, the state of the system collapses to
\vspace{-.35cm}
\begin{eqnarray}
&& \ket\phi=\frac{\alpha cos\theta\ket0+\beta\ket1}{\sqrt{|\alpha|^2cos^2\theta+|\beta|^2}}\nonumber\\
&& \approx (\alpha cos\theta\ket0+\beta\ket1)(1-|\alpha|^2\theta^2)^{-\frac{1}{2}}\nonumber\\
&& \approx (\alpha(1-\frac{\theta^2}{2})\ket0+\beta\ket1)(1+\frac{1}{2}|\alpha|^2\theta^2)\nonumber\\
&& \label{qlock_7}= \alpha(1-\frac{|\beta|^2\theta^2}{2})\ket0+\beta(1+\frac{|\alpha|^2\theta^2}{2})\ket1
\end{eqnarray}

Evidently, after one iteration, the state of the system is perturbed. Denoting $\alpha'=\alpha(1-\frac{|\beta|^2\theta^2}{2})$ and $\beta'=\beta(1+\frac{|\alpha|^2\theta^2}{2})$, it is to be noted that $|\alpha'|<|\alpha|$, while $|\beta'|>|\beta|$, provided $\alpha\ne0$ and $\beta\ne0$. The state of the system is weakly perturbed towards the \textit{z}-basis eigen state $|1\rangle$. If we repeat the unitary transformation $U$ on the composite system, followed by a measurement on the ancillary qubit, the probability of obtaining the ancilla in the state $\ket1$, will be $p'_1\approx|\alpha'|^2\theta^2$, which is less than $p_1$. Hence, the state of the ancilla collapses to the state $\ket0$ with an even higher probability than the previous one. In this case, the state of the system $\ket\phi$ is further perturbed towards the state $\ket1$. In other words, the co-efficient of the eigen state $\ket1$ in the superposition state of the system increases slightly in magnitude. If we repeat this protocol a large number of times (while assuming that the state $\ket0$ is obtained in each measurement), we will obtain the final state $\ket\phi$ of the system (initially in the state $\alpha\ket0+\beta\ket1$) arbitrarily close to the \textit{z}-basis eigen state $\ket1$. Our assumption, that the state $\ket0$ is obtained after every measurement of the ancilla, is justifiable since $\theta$ can take any arbitrary small value ($\theta\rightarrow0$), such that the probability of getting $\ket1$ is also extremely small (having a $\theta^2$ dependence). Additionally, our assumption is greatly aided by the fact that the probability of obtaining the state $\ket1$ decreases with each iteration, since it has a $|\alpha|^2$ dependence, and $|\alpha|$ decreases with each iteration.

It is evident that, after each iteration of the verification box, the state of the system does not change if it is initially in any of the $z$-basis eigenstates \{$\ket{0}$,$\ket1$\}. It can also be pointed out that the eigen states of the \textit{z}-basis act as ``fixed points" for the evolution protocol mentioned above, where $\ket0$ and $\ket1$ behave as unstable and stable fixed points respectively. However, if a generic superposition state $\alpha\ket0+\beta\ket1$ ($\alpha,\beta\ne0$) is fed into our verification box, then the state of the system asymptotically approaches the stable fixed point $\ket1$. In this way, we can distinguish the eigenstate $\ket0$ from any arbitrary superposition state $\alpha\ket0+\beta\ket1$.

A single iteration of the standard Grover's algorithm rotates any arbitrary single qubit state $\ket\psi$ through an angle of $\frac{\pi}{2}$ about the \textit{z}-axis, implying that the state transforms back to itself periodically after 4 iterations. Our verification algorithm uses an ancillary qubit, which is weakly coupled to the system qubit. It involves a repetitive series of coupling and decoupling of the ancilla (environment) with the system. Hence, some information about the state of the system is leaked to the environment. During the process, the coupling between the two qubits is followed by measurement of the ancilla in \textit{z}-basis, which results in collapsing the state of the ancilla to a \textit{z}-basis eigenstate ($\ket0$) with high probability. Correspondingly, the state of the system is only weakly perturbed, towards the other eigenstate ($\ket1$). This results in decoupling the system from the environment. The same procedure for coupling and decoupling is repeated a large number of times. 

The coupling process takes place by means of the unitary transformation $U\equiv Controlled^0-R_x(2\theta)$. To preserve the eigenstate $\ket0$ and transform any other arbitrary state to the eigenstate $\ket1$, the $Controlled^0-R_x(2\theta)$ operation is used. In a way, it prevents any information about the state of the system related to the eigenstate $\ket0$ from passing to the environment (ancilla). The operation $U$ is also equivalent to the operation $[R_z\otimes I]e^{-iC^0NOT_{12}\theta}$, a \textit{conditional} ``decay" operation, which acts as an eigenstate amplifier, amplifying the eigenstate $\ket1$ in the superposition state $\ket\psi$.

\textbf{Quantum Locker: Based on a Quantum OTP \label{qlock_III}} 

\textbf{Definition of a Quantum Locker and a Quantum OTP}

Informally, the proposed scheme of a quantum locker consists of three stages,
\begin{itemize}
\item \textbf{First stage,} where a message and certain parameters, required for the verification of One Time Password (OTP), are fed into the locker. 
\item \textbf{Second stage,} where a quantum OTP state is teleported to the intended receiver. 
\item \textbf{Third stage,} where a protocol is presented for the verification of the OTP and transfer of the message.
\end{itemize}

Ideally, an OTP is expected to have the following properties,
\begin{itemize}
\item \textbf{Verifiability,} i.e., it can be verified by the locker.
\item \textbf{Unforgeability,} i.e., an OTP can neither be counterfeited nor can it be used more than once to access the message stored in the locker. 
\end{itemize}

\textbf{The Quantum Locker Scheme}

For the purpose of brevity, two parties, Alice and Bob, are introduced to describe the scheme.~Initially, they share a maximally entangled pair of qubits in the state, $\frac{\ket{00}+\ket{11}}{\sqrt{2}}$, which is to be used as the teleportation channel between the two. Alice stores a message along with an OTP in the locker, situated at a secure location accessible to both parties, so that Bob could access the message in the future. She then teleports the OTP to Bob, by using which he verifies his authenticity and retrieves the message from the locker.

\textbf{Locker:} It consists of $m$ qubits which store the message (named as message qubits), and one ancillary qubit, which are inaccessible to any outside party. The message qubits can be either $|0\rangle$ or $|1\rangle$. The $m$ qubit message is, therefore, an arbitrary string composed of $|0\rangle$s and $|1\rangle$s only. Let $\Ket{a^{(i)}_L}$ and $\Ket{\delta_L}$ denote the $i^{th}$ message qubit and the ancillary qubit, in the locker respectively. The locker has an input slot through which any party may input the password qubit, as well as $m$ other input slots into which the party may place $m$ qubits in $\ket0$ state, meant for retrieving the message. In the whole scheme, it is assumed that Bob has the prior knowledge about the basis of the stored message qubits. 

\textbf{The Protocol}

\textbf{Storage of Message And Password}

The $m$ message qubits in the locker are initially prepared in $\Ket0$ state. Now, Alice encodes her message in the locker by applying X gates on the requisite message qubits, in order to make a binary string of $|0\rangle$s and $|1\rangle$s. This feature is provided by the locker itself. For reasons that have been stated in a later subsection, a string with all qubits in $|0\rangle$ state is not a valid message. Henceforth, for password purposes, Alice must randomly choose and input the values of $\theta_1$, $\theta_2$ and $\theta_3$ to the locker, where $\theta_1$, $\theta_2$, $\theta_3$  $\epsilon$ $[-\pi, \pi]$. These values act as parameters for the rotation operator denoted by, $R^{-1}(\theta_1,\theta_2,\theta_3)$, where $R^{-1}(\theta_1,\theta_2,\theta_3)=R_x(-\theta_1)R_y(-\theta_2)R_z(-\theta_3)$.  

\textbf{Generation of OTP State}

Alice owns an ancilla qubit which is initially kept in $\ket0$ state. She is equipped with a portable device which can perform the operation, $R(\theta_1,\theta_2,\theta_3)=R_z(\theta_3)R_y(\theta_2)R_x(\theta_1)$, where $\theta_1$, $\theta_2$ and $\theta_3$ are the same parameters previously used as inputs to the locker. Since this device is portable, she can perform this operation anytime and anywhere at her will. Alice generates the OTP state denoted by, $\ket\psi$ by performing the above operation on her ancilla qubit, and sends it to Bob through the teleportation channel. 

\textbf{Verification of OTP}

For retrieving the message, Bob must store the teleported state, $\ket\psi$ in a qubit. The locker implements the operation $R^{-1}(\theta_1,\theta_2,\theta_3)$ on this qubit to create a new state, represented by $\ket\phi$. He also keeps $m$ qubits (named as blank qubits) possessing $|0\rangle$ state in the specified slots of the locker. For the purpose of verification, a verification box $V$ has been designed, the detailed working of which has already been discussed. The verification box $V$ acts on $\ket\phi$ and produces two distinguishable results depending on whether the qubit entered by some party is in the state $\ket\psi$ (the state originally prepared by Alice) or not.

\textbf{CASE I - Entering Correct Password:}
In this case, the password entered into the locker is in the state $\ket\psi$. The operation of $R^{-1}(\theta_1,\theta_2,\theta_3)$ on $\ket\psi$ yields $\ket\phi\equiv\ket0$. The state does not change by applying the verification box $V$ on $\ket\phi$ a large number of times.

\textbf{CASE II - Entering Wrong Password:\label{qlock_Sub.1}}
In this case, the password entered into the locker is in a state $|\psi'\rangle$, which is different from the state $\ket\psi$. The operation of $R^{-1}(\theta_1,\theta_2,\theta_3)$ on $|\psi'\rangle$ yields $\ket\phi\equiv\alpha\ket0+\beta\ket1$, where $\alpha$ and $\beta$ are arbitrary complex numbers. In this case, after the operation of the verification box $V$ a large number of times, the state $\ket\phi$ is transformed into the state $\ket1$, which can be distinguished from the state $\ket0$.

\textbf{Transfer of Message \label{qlock_Sub.2}}

\begin{figure}[!ht]
\centering
\includegraphics[scale=2.0]{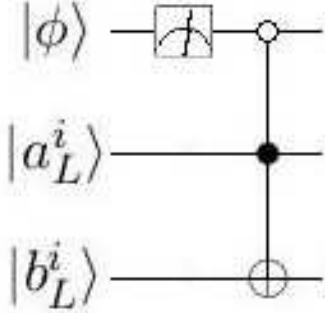}
\caption{\emph{A C$^{2}$NOT gate is used for transferring information from the $i^{th}$ message qubit $\Ket{a^{(i)}_L}$ to the $i^{th}$ blank qubit $\Ket{b^{(i)}_L}$. This transfer of information takes place only if the qubit $\ket\phi$ is found to be in the state $|0\rangle$ after the measurement.}}
\label{qlock_Figure4}
\end{figure}

After the operation of the verification box $V$ on $\ket\phi$, a measurement is performed on this qubit (in \textit{z}-basis). To transfer the $i^{th}$ message qubit $\Ket{a^{(i)}_L}$ to the $i^{th}$ blank qubit $\Ket{b^{(i)}_L}$, we use the circuit shown in Fig.~\ref{qlock_Figure4}.

There are $m$ such Toffoli gates as used in Fig.~\ref{qlock_Figure4}, where the first control qubit is the same ($\ket{\phi}$, following measurement) for all gates, while the second control and the target qubits of the $i$th Toffoli gate are in $\Ket{a^{(i)}_L}$ and $\Ket{b^{(i)}_L}$ states respectively. It can be easily seen that the measurement of $\ket\phi$ gives the outcome $\ket0$ when the entered password is correct. Hence, $\Ket{b^{(i)}_L}$ becomes $\ket1$ if $\Ket{a^{(i)}_L}$ is in the state $\ket1$, otherwise it remains in the state $\ket0$. Thus, the information stored in the message qubits is transferred to the blank qubits entered into the locker by the party, which are now ready for retrieval.

For the case when the entered password is incorrect, measurement of $\ket\phi$ following the verification box gives the outcome $\ket1$ with an arbitrarily high probability. Hence, none of the $\Ket{b^{(i)}_L}$ is flipped, resulting in no transfer of information. This explains why $\Ket{a^{(i)}_L}=\ket0\forall i$ is not a valid message- we consider this case as ``informationless".

\section{Methods}

\textbf{Implementation in IBM Quantum Computer \label{qlock_experiment}}

\textbf{Measurement of the ancillary qubit following each iteration}

\begin{figure}[!ht]
\centering
\includegraphics[width=\linewidth]{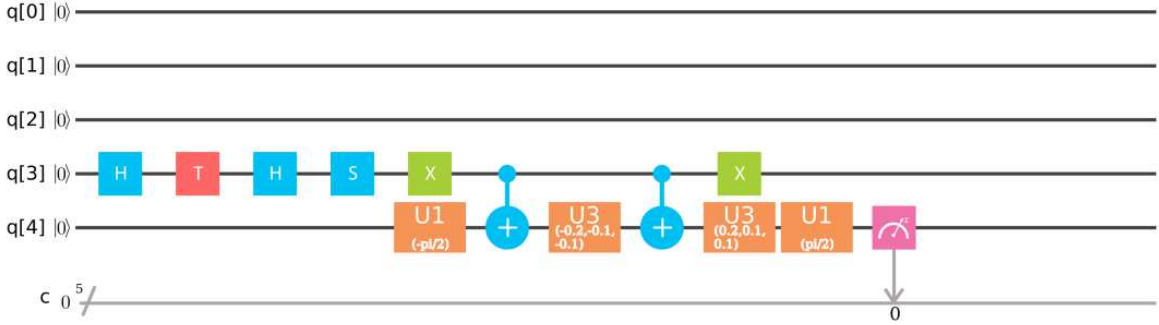}
\caption{\emph{IBM quantum circuit illustrating a single iteration of the Verification Box $V$ with an initial system state $|\phi\rangle \equiv cos\frac{\pi}{8}|0\rangle+sin\frac{\pi}{8}|1\rangle$ and ancilla state $\ket{\delta}\equiv|0\rangle$.}}
\label{qlock_Figure5}
\end{figure}

In the circuit shown in Fig.~\ref{qlock_Figure5}, measurement of the ancillary qubit is performed in \textit{z}-basis. To obtain the experimental density matrix for state tomography, we must perform measurements on the ancilla in the \textit{x} and \textit{y} bases as well. By applying $H$ or $S^{\dagger}H$ gate(s)  before the measurement operation, the ancillary qubit can be measured in the $x$ or $y$ basis respectively. The experimental results are provided in Table~\ref{qlock_tab2}.
\begin{table}[H]
\centering
\caption{\emph{Table depicting the experimental results for the probability outcome of the ancilla state ($\Ket{\delta}$).}}
\begin{tabular}{ccc}
\hline
\hline
\multicolumn{3}{c}{Measurement of Ancilla State (For 8192 shots)}\\
\hline
Basis & Probability of $\ket{0}$ & Probability of $\ket{1}$\\
\hline
\textit{x}  & 0.498 & 0.502 \\
\textit{y}  & 0.710 & 0.290 \\
\textit{z}  & 0.938 & 0.063 \\ 
\hline
\hline
\end{tabular}
\label{qlock_tab2}
\end{table}

To check the accuracy of our experimental results, quantum state tomography is performed. The theoretical density matrix of the ancilla is given by,
\begin{equation}
    \rho^{T}=p_0\ket{0}\bra{0}+p_1\ket{1}\bra{1}
\end{equation}

In the experiment, we have chosen $\theta=0.2$ (Eq.~\eqref{qlock_1}). Also, the initial state of the system (given by $\ket{\phi}$) is taken such that $\alpha=cos\frac{\pi}{8}$ and $\beta=sin\frac{\pi}{8}$. Putting the values of $\alpha$, $\beta$ and $\theta$, and obtaining $p_0$ and $p_1$ (Eqs.~\eqref{qlock_5},\eqref{qlock_6}), we evaluate $\rho^T$ to be
\begin{equation}
    \rho^T= \left[ {\begin{array}{cc}
        0.966 & 0.000 \\
        0.000 & 0.034 \\
    \end{array} }\right]
\end{equation}

The following equation gives the experimental density matrix for a single qubit.
\begin{equation}
    \rho^E= \frac{1}{2}\left[I+<x>\sigma_x+<y>\sigma_y+<z>\sigma_z\right]
\end{equation}

Here $<x>$, $<y>$ and $<z>$ are related to the experimental outcomes of projective measurements in the \textit{x}, \textit{y} and \textit{z} bases respectively, and are known as Stokes parameters. For a given basis $j$, its corresponding Stokes parameter is given by $<j>=p_{\ket{0_j}}-p_{\ket{1_j}}$, where $\ket{0_j}$ and $\ket{1_j}$ are eigenstates of the given basis.
\begin{equation}
  \rho^{E}_{q}=
  \left[ {\begin{array}{cc}
   0.937 & -0.002 \\
   -0.002 & 0.063 \\
  \end{array} } \right]+ i\left[ {\begin{array}{cc}
   0.000 & -0.210 \\
   0.210 & 0.000 \\
  \end{array} } \right]
\end{equation}

\begin{figure}[H]
\centering
\includegraphics[width=\linewidth]{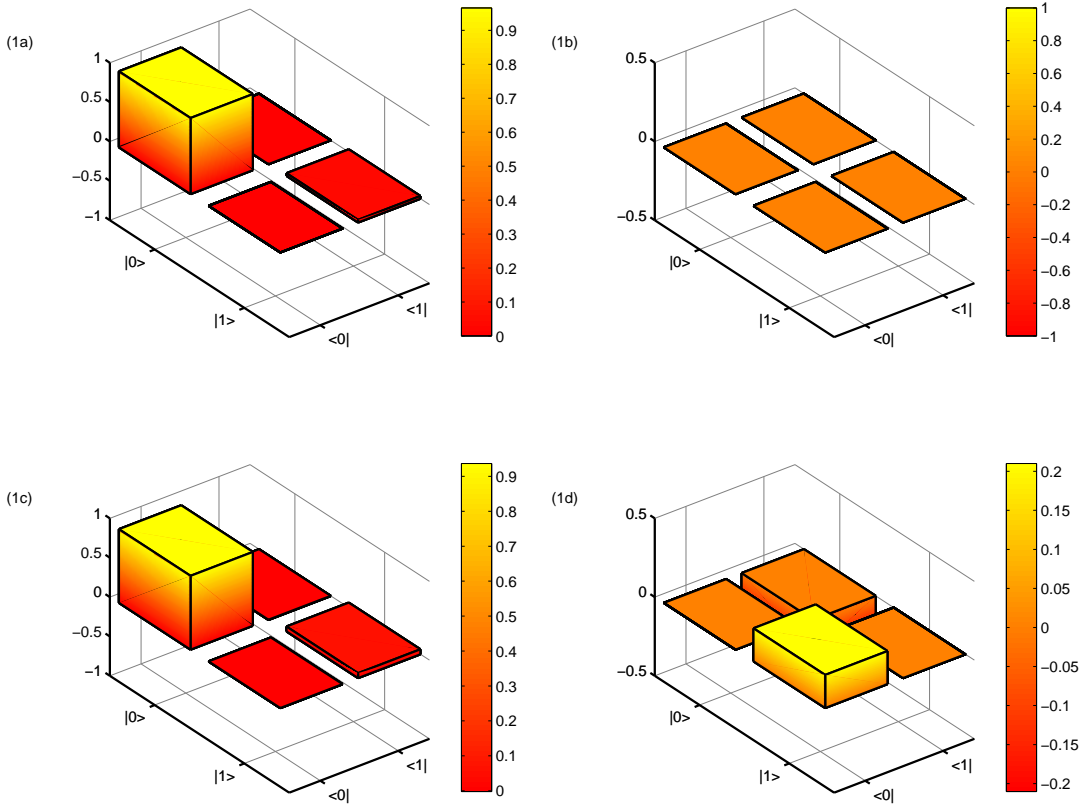}
    \caption{\textbf{Measurement of Ancillary Qubit}: \emph{Real (left) and imaginary (right) parts of the reconstructed theoretical (1a,1b) and experimental (1c,1d) density matrices for the ancilla state.}}
    \label{qlock_Figure6}
\end{figure}

Fig.~\ref{qlock_Figure6} compares the theoretical and experimental density matrices of ancilla state. It is evident that the real part of the experimental density matrix is in good agreement with the theoretical one. Hence, it can be concluded that the ancilla state collapses to the eigenstate $\ket{0}$ with a high probability.

\textbf{Evolution of the system qubit after several iterations}

\begin{figure}[!ht]
\centering
\includegraphics[width=\linewidth]{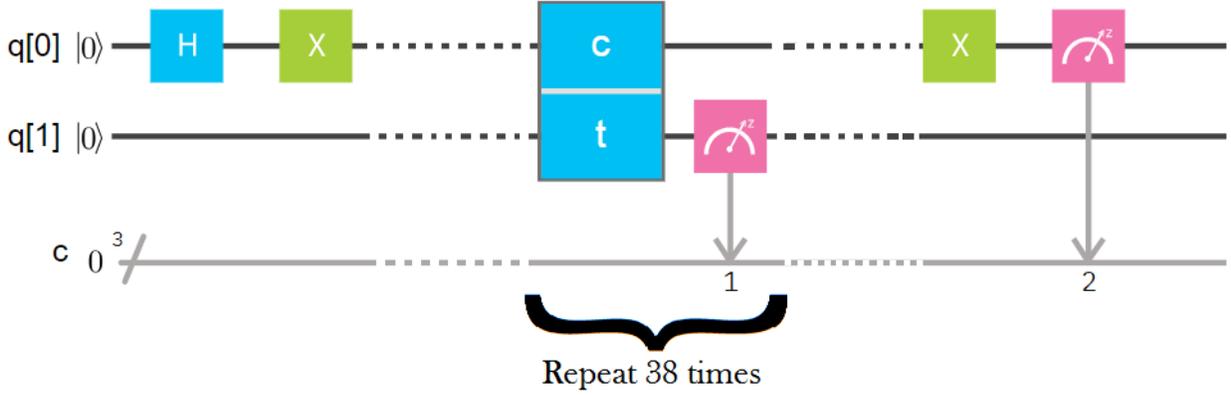}
\caption{\emph{Circuit clearly depicting the working of Verification Box $V$ with an initial system state $|+\rangle$ and ancilla state $|0\rangle$.}}
\label{qlock_Figure7} 
\end{figure}

We have used IBM's Custom Topology to simulate the working of our verification algorithm, where $38$ iterations are taken into consideration.~We use this model to examine the evolution of the system qubit $\ket{\phi}$ (initially in the state $\ket{+}$). In the above circuit (Fig.~\ref{qlock_Figure7}), q[0] and q[1] denote the system and the ancillary qubits respectively. The two qubit gates (c-t) represent the $Controlled-R_x(2\theta)$ operation, where $\theta=0.1$. It is to be noted that, in our verification algorithm, we use the $Controlled^0-R_x(2\theta)$ operation, which can be equivalently produced by applying $\sigma_x$ gates on system qubit (q[0]) before and after the $Controlled-R_x(2\theta)$ operation.

After simulating the above quantum circuit (Fig.~\ref{qlock_Figure7}) for $8192$ shots, $248$ different measurement outcomes are obtained, from which only two outcomes are found to occur a large number of times ($6836$ times). Specifically, each of these two outcomes are such that, all the $38$ ancilla measurements (following every iteration) yield the state $\ket{0}$. Out of $6836$ times, the system qubit measurement yields the state $\ket{1}$ $4116$ times and the state $\ket{0}$ $2720$ times. Hence, the system qubit is ultimately found to be in the eigenstate $\ket{1}$ with a high probability, even though it was initially prepared in the equal superposition state $\ket{+}$, as predicted by our proposed algorithm.

\section{Discussion \label{qlock_discussion}}

Here, we have discussed some significant features of the proposed quantum locker. It is also explained, how a quantum locker based on the principles of quantum mechanics offers significantly more security than a classical one which we use in our daily life.

\textbf{Security Aspects}

The security of the protocol is guaranteed by the secrecy of the parameters $\theta_1$, $\theta_2$ and $\theta_3$, which are arbitrary real numbers chosen by the sender. It is in practice impossible to guess any arbitrarily chosen real number from a given interval $[a,b]$ which is dense in real numbers. Even for a single qubit OTP, like the case discussed so far in our paper, there are $3$ real parameters, each one in the interval $[-\pi,\pi]$, which have to be guessed exactly in order to produce the correct password and retrieve the secured message. In a classical OTP, each of its characters is chosen from a finite character set, for instance, the ASCII character set, which makes it, in practice, ``crackable" with some finite probability. Thus, a classical OTP, even with multiple characters/digits, is no match for a quantum one consuming just a single qubit.

A profound feature of the protocol discussed here is the inability of a purported eavesdropper to steal the OTP, which Alice (the message encoder) teleports to Bob (the intended retriever). The members (qubits) of the EPR pair which serve as the teleportation channel is shared by the two parties only. Each operation required in the teleportation protocol is performed locally by either Alice or Bob.~The only part of the protocol which is vulnerable to an eavesdropper is the series of classical channels used in the end. However, no productive information could possibly be obtained by stealing any data transmitted through these classical links. Unlike this, classical OTP must be communicated to the intended retriever through a classical link, which makes it vulnerable to an outside party.

Another unique feature of the presented protocol is the intended retriever's ignorance of the OTP necessary for retrieval of the stored message. In our discussion, we have imagined that a person named Bob is the intended retriever of some message encoded by Alice. Bob receives the arbitrary state prepared by Alice by means of the teleportation protocol, which plays the role of the OTP. He has absolutely no idea as to which random values were chosen by Alice for the parameters $\theta_1$, $\theta_2$ and $\theta_3$ to prepare the OTP. In other words, no outside party (not even the one intended to retrieve the message) can forge the OTP. The necessary parameters for producing the correct password is known only to the encoder of the message.

\begin{figure}[!ht]
\centering
\includegraphics[scale=1.2]{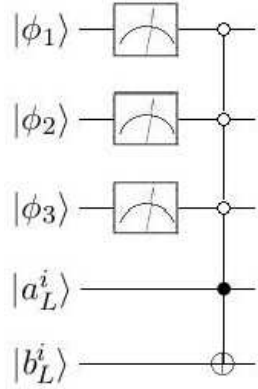}
\caption{\emph{A C$^4$NOT gate is used for transferring information from the $i^{th}$ message qubit $\Ket{a^{(i)}_L}$ to the $i^{th}$ blank qubit $\Ket{b^{(i)}_L}$, in case where the OTP comprises 3 qubits. Hence, the entered password must also consist of 3 qubits, whose states are transformed to the states $\ket{\phi_1}$, $\ket{\phi_2}$ and $\ket{\phi_3}$ respectively.}}
\label{qlock_Figure8}
\end{figure}

In our protocol, the OTP is composed of a single qubit. The security can be further enhanced by the use of multiple qubits, say $n$, for the purpose of generating the OTP, where the values of the parameters ($\theta_1$, $\theta_2$ and $\theta_3$) are chosen independently for each qubit. We follow the same protocol as discussed here. All the $n$ qubits are teleported to the intended retriever (say Bob) simultaneously. Bob places these qubits in their specified slots in the locker (along with the blank qubits for message retrieval). The verification protocol acts simultaneously on the OTP qubits (considering $n$ in-built Verification Boxes). The transfer of message takes place if and only if all the $n$ qubits comprising the OTP collapse to the state $\ket0$ on measurement (similar to the case discussed earlier). In Section~\ref{qlock_Sub.2}, the Toffoli gate, viz. the $C^2NOT$ gate, the case in which $n=1$, was used for message retrieval. In general, one may use a $C^{n+1}NOT$ gate for n+1 qubit system. Note that, a simple case of $n=3$ has been shown in Fig. \ref{qlock_Figure8}.

\textbf{Limitations}

The quantum locker scheme discussed here allows Bob to retrieve the message stored in the locker with certainty, viz. if the entered password is correct. When a wrong password is entered, then the protocol ensures that the message is not transferred to the blank qubits. However, the verification algorithm presented in Section \ref{qlock_veri} being probabilistic, there is always a minuscule possibility that the message is still transferred to the blank qubits. Consider the Case II stated earlier in Section~\ref{qlock_Sub.1}, namely the password entered into the locker is in a state $|\psi'\rangle$, implying that $\ket\phi\equiv\alpha\ket0+\beta\ket1$ ($\beta\ne0$). When $V$ operates on $\ket\phi$ there is an extremely slim chance of obtaining $\ket1$ whenever a projective measurement is performed on the ancillary $\Ket{\delta_L}$ (Check Eq.(\ref{qlock_6})). In such a situation, the $\ket\phi$ collapses to the state $\ket0$, and remains in this state in subsequent iterations. Thus, the message will be transferred to the blank qubits in such a scenario.

\section{Conclusion \label{qlock_conclusion}}

To conclude, we have proposed here a verification algorithm, in a way that it takes an unequal superposition state and transforms it into an eigenstate. The algorithm also \textit{distinguishes} any arbitrary superposition state from an eigenstate. We have introduced the concept of a locker based on the principles of quantum mechanics, which takes an arbitrarily generated quantum state as a password. The locker subsequently destroys the quantum information encoded in the password qubit(s) through the course of its operation. Hence, the password acts as a ``One-Time Password". We have presented a password verification protocol based on the principles of weak measurement. The working of the verification algorithm has been experimentally realized with a high fidelity. 

In our discussed protocol, we have imagined the contents of the locker to be some form of a message encoded by the sender, composed of a certain number of qubits. The basic principles of our quantum locker scheme could be extended to various other applications. Instead of storing a message, one may consider storing some physical entity, as one may wish, which could be accessible to the intended retriever once the entered password has been verified by the locker.

\providecommand{\WileyBibTextsc}{}
\let\textsc\WileyBibTextsc
\providecommand{\othercit}{}
\providecommand{\jr}[1]{#1}
\providecommand{\etal}{~et~al.}

\section*{Acknowledgments}
\label{qlock_acknowledgments}
A.D. would like to thank Kishore Vaigyanik Protsahan Yojana (KVPY) for providing financial support in the undertaking of this paper. B.K.B. acknowledges the support of Inspire Fellowship awarded by DST, Government of India. S.R. would like to thank IISER Kolkata for providing hospitality during which this work has been done. The authors acknowledge the support of IBM Quantum Experience for producing experimental results. The views expressed are those of the authors and do not reflect the official policy or position of IBM or the IBM Quantum Experience team.
\section*{Author contributions}
A.D., S.R. and B.K.B. have discussed and designed all the quantum circuits, and have written the paper. B.K.B. and A.D. performed all the experiments in IBM Quantum Experience platform. P.K.P. has thoroughly checked and reviewed the manuscript.  

\section*{Competing interests}
The authors declare no competing financial as well as non-financial interests. 


\begin{thebibliography}{[10]}

\bibitem{219067:4918030}
\bibinfo{author}{Grover, L.~K.}
\newblock \bibinfo{title}{A fast quantum mechanical algorithm for database
  search}.
\newblock In \emph{\bibinfo{booktitle}{Proceedings of the twenty-eighth annual
  ACM symposium on Theory of computing}}, \bibinfo{pages}{212--219}
  (\bibinfo{year}{1996}).

\bibitem{219067:4918149}
\bibinfo{author}{Biron, D.}, \bibinfo{author}{Biham, O.},
  \bibinfo{author}{Biham, E.}, \bibinfo{author}{Grassl, M.} \&
  \bibinfo{author}{Lidar, D.~A.}
\newblock \bibinfo{title}{Generalized grover search algorithm for arbitrary
  initial amplitude distribution} (\bibinfo{year}{1999}).
 
\bibitem{219067:4918245}
\bibinfo{author}{Biham, E.} \emph{et~al.}
\newblock \bibinfo{title}{Analysis of generalized grover quantum search
  algorithms using recursion equations}.
\newblock \emph{\bibinfo{journal}{Phys. Rev. A}}
  \textbf{\bibinfo{volume}{63}}, \bibinfo{pages}{012310}
  (\bibinfo{year}{2000}).

\bibitem{219067:4918839}
\bibinfo{author}{H\ensuremath{\varnothing }yer, P.}
\newblock \bibinfo{title}{Arbitrary phases in quantum amplitude amplification}.
\newblock \emph{\bibinfo{journal}{Phys. Rev. A}}
  \textbf{\bibinfo{volume}{62}}, \bibinfo{pages}{052304}
  (\bibinfo{year}{2000}).
  
\bibitem{219067:4918881}
\bibinfo{author}{Long, G.-L.}
\newblock \bibinfo{title}{Grover algorithm with zero theoretical failure rate}.
\newblock \emph{\bibinfo{journal}{Phys. Rev. A}}
  \textbf{\bibinfo{volume}{64}}, \bibinfo{pages}{022307}
  (\bibinfo{year}{2001}).

\bibitem{219067:4918978}
\bibinfo{author}{Long, G.~L.}, \bibinfo{author}{Li, Y.~S.},
  \bibinfo{author}{Zhang, W.~L.} \& \bibinfo{author}{Niu, L.}
\newblock \bibinfo{title}{Phase matching in quantum searching}.
\newblock \emph{\bibinfo{journal}{Phys. Lett. A}}
  \textbf{\bibinfo{volume}{262}}, \bibinfo{pages}{27--34}
  (\bibinfo{year}{1999}).

\bibitem{219067:4919007}
\bibinfo{author}{Long, G.-L.}, \bibinfo{author}{Li, X.} \&
  \bibinfo{author}{Sun, Y.}
\newblock \bibinfo{title}{Phase matching condition for quantum search with a
  generalized initial state}.
\newblock \emph{\bibinfo{journal}{Phys. Lett. A}}
  \textbf{\bibinfo{volume}{294}}, \bibinfo{pages}{143--152}
  (\bibinfo{year}{2002}).
  
\bibitem{219067:4919049}
\bibinfo{author}{Liu, Y.}
\newblock \bibinfo{title}{An exact quantum search algorithm with arbitrary
  database}.
\newblock \emph{\bibinfo{journal}{Int. J. Theor. Phys.}} \textbf{\bibinfo{volume}{53}}, \bibinfo{pages}{2571--2578}
  (\bibinfo{year}{2014}).


\bibitem{219067:4919155}
\bibinfo{author}{Brassard, G.}, \bibinfo{author}{H\ensuremath{\varnothing }yer,
  P.} \& \bibinfo{author}{Tapp, A.}
\newblock \bibinfo{title}{Quantum counting}.
\newblock In \emph{\bibinfo{booktitle}{International Colloquium on Automata,
  Languages, and Programming}}, \bibinfo{pages}{820--831}
  (\bibinfo{year}{1998}).

\bibitem{219067:4919157}
\bibinfo{author}{Mosca, M.} \emph{et~al.}
\newblock \bibinfo{title}{Quantum searching, counting and amplitude
  amplification by eigenvector analysis}.
\newblock In \emph{\bibinfo{booktitle}{MFCS'98 workshop on Randomized
  Algorithms}}, \bibinfo{pages}{90--100} (\bibinfo{year}{1998}).

\bibitem{219067:4919441}
\bibinfo{author}{Biham, E.} \& \bibinfo{author}{Kenigsberg, D.}
\newblock \bibinfo{title}{Grover's quantum search algorithm for an arbitrary
  initial mixed state}.
\newblock \emph{\bibinfo{journal}{Phys. Rev. A}}
  \textbf{\bibinfo{volume}{66}}, \bibinfo{pages}{062301}
  (\bibinfo{year}{2002}).

\bibitem{219067:4929851}
\bibinfo{author}{Lamport, L.}
\newblock \bibinfo{title}{Password authentication with insecure communication}.
\newblock \emph{\bibinfo{journal}{Commun. ACM}} \textbf{\bibinfo{volume}{24}},
  \bibinfo{pages}{770--772} (\bibinfo{year}{1981}).

\bibitem{219067:4929852}
\bibinfo{author}{Bellovin, S.~M.} \& \bibinfo{author}{Merritt, M.}
\newblock \bibinfo{title}{Limitations of the kerberos authentication system}.
\newblock \emph{\bibinfo{journal}{ACM SIGCOMM Comput. Commun. Rev.}}
  \textbf{\bibinfo{volume}{20}}, \bibinfo{pages}{119--132}
  (\bibinfo{year}{1990}).

\bibitem{219067:4929853}
\bibinfo{author}{Haller, N.}, \bibinfo{author}{Metz, C.},
  \bibinfo{author}{Nesser, P.} \& \bibinfo{author}{Straw, M.}
\newblock \bibinfo{title}{A one-time password system} (\bibinfo{year}{1998}).

\bibitem{219067:4919523}
\bibinfo{author}{Wootters, W.~K.} \& \bibinfo{author}{Zurek, W.~H.}
\newblock \bibinfo{title}{A single quantum cannot be cloned}.
\newblock \emph{\bibinfo{journal}{Nature}} \textbf{\bibinfo{volume}{299}},
  \bibinfo{pages}{802--803} (\bibinfo{year}{1982}).

\bibitem{219067:4919565}
\bibinfo{author}{Lo, H.-K.} \& \bibinfo{author}{Chau, H.~F.}
\newblock \bibinfo{title}{Unconditional security of quantum key distribution
  over arbitrarily long distances}.
\newblock \emph{\bibinfo{journal}{Science}} \textbf{\bibinfo{volume}{283}},
  \bibinfo{pages}{2050--2056} (\bibinfo{year}{1999}).

\bibitem{219067:4919569}
\bibinfo{author}{Mayers, D.}
\newblock \bibinfo{title}{Unconditional security in quantum cryptography}.
\newblock \emph{\bibinfo{journal}{JACM}}
  \textbf{\bibinfo{volume}{48}}, \bibinfo{pages}{351--406}
  (\bibinfo{year}{2001}).

\bibitem{219067:4919611}
\bibinfo{author}{Mayers, D.} \& \bibinfo{author}{Yao, A.}
\newblock \bibinfo{title}{Quantum cryptography with imperfect apparatus}.
\newblock In \emph{\bibinfo{booktitle}{Foundations of Computer Science, 1998.
  Proceedings. 39th Annual Symposium on}}, \bibinfo{pages}{503--509}
  (\bibinfo{year}{1998}).

\bibitem{219067:4919614}
\bibinfo{author}{Biham, E.}, \bibinfo{author}{Boyer, M.},
  \bibinfo{author}{Boykin, P.~O.}, \bibinfo{author}{Mor, T.} \&
  \bibinfo{author}{Roychowdhury, V.}
\newblock \bibinfo{title}{A proof of the security of quantum key distribution
  (extended abstract)}.
\newblock In \emph{\bibinfo{booktitle}{Proceedings of the Thirty-second Annual
  ACM Symposium on Theory of Computing}}, \bibinfo{pages}{715--724}
  (\bibinfo{publisher}{ACM}, \bibinfo{address}{New York, NY, USA},
  \bibinfo{year}{2000}).

\bibitem{219067:4919615}
\bibinfo{author}{Shor, P.~W.} \& \bibinfo{author}{Preskill, J.}
\newblock \bibinfo{title}{Simple proof of security of the bb84 quantum key
  distribution protocol}.
\newblock \emph{\bibinfo{journal}{Phys. rev. lett.}}
  \textbf{\bibinfo{volume}{85}}, \bibinfo{pages}{441} (\bibinfo{year}{2000}).

\bibitem{219067:4921045}
\bibinfo{author}{Fujiwara, M.} \emph{et~al.}
\newblock \bibinfo{title}{Unbreakable distributed storage with quantum key
  distribution network and password-authenticated secret sharing}.
\newblock \emph{\bibinfo{journal}{Sci. Rep.}}
  \textbf{\bibinfo{volume}{6}}, \bibinfo{pages}{28988} (\bibinfo{year}{2016}).

\bibitem{219067:4929854}
\bibinfo{author}{Horodecki, R.}, \bibinfo{author}{Horodecki, P.},
  \bibinfo{author}{Horodecki, M.} \& \bibinfo{author}{Horodecki, K.}
\newblock \bibinfo{title}{Quantum entanglement}.
\newblock \emph{\bibinfo{journal}{Rev. Mod. Phys.}}
  \textbf{\bibinfo{volume}{81}}, \bibinfo{pages}{865} (\bibinfo{year}{2009}).

\bibitem{219067:4921046}
\bibinfo{author}{Ekert, A.~K.}
\newblock \bibinfo{title}{Quantum cryptography based on bell's theorem}.
\newblock \emph{\bibinfo{journal}{Phys. Rev. Lett.}}
  \textbf{\bibinfo{volume}{67}}, \bibinfo{pages}{661} (\bibinfo{year}{1991}).

\bibitem{219067:4921047}
\bibinfo{author}{Bennett, C.~H.}
\newblock \bibinfo{title}{Quantum cryptography using any two nonorthogonal
  states}.
\newblock \emph{\bibinfo{journal}{Phys. Rev. Lett.}}
  \textbf{\bibinfo{volume}{68}}, \bibinfo{pages}{3121} (\bibinfo{year}{1992}).

\bibitem{219067:4921048}
\bibinfo{author}{Du\v{s}ek, M.}, \bibinfo{author}{Haderka, O.},
  \bibinfo{author}{Hendrych, M.} \& \bibinfo{author}{My\v{s}ka, R.}
\newblock \bibinfo{title}{Quantum identification system}.
\newblock \emph{\bibinfo{journal}{Phys. Rev. A}}
  \textbf{\bibinfo{volume}{60}}, \bibinfo{pages}{149} (\bibinfo{year}{1999}).

\bibitem{219067:4921049}
\bibinfo{author}{Barnum, H.~N.}
\newblock \bibinfo{title}{Quantum secure identification using entanglement and
  catalysis}.
\newblock \emph{\bibinfo{journal}{arXiv preprint quant-ph/9910072}}
  (\bibinfo{year}{1999}).

\bibitem{219067:4921050}
\bibinfo{author}{Jensen, J.~G.} \& \bibinfo{author}{Schack, R.}
\newblock \bibinfo{title}{Quantum authentication and key distribution using
  catalysis}.
\newblock \emph{\bibinfo{journal}{arXiv preprint quant-ph/0003104}}
  (\bibinfo{year}{2000}).

\bibitem{219067:4921051}
\bibinfo{author}{Gottesman, D.} \& \bibinfo{author}{Chuang, I.}
\newblock \bibinfo{title}{Quantum digital signatures}.
\newblock \emph{\bibinfo{journal}{arXiv preprint quant-ph/0105032}}
  (\bibinfo{year}{2001}).

\bibitem{219067:4921052}
\bibinfo{author}{Moulick, S.~R.} \& \bibinfo{author}{Panigrahi, P.~K.}
\newblock \bibinfo{title}{Quantum cheques}.
\newblock \emph{\bibinfo{journal}{Quantum Inf. Process.}}
  \textbf{\bibinfo{volume}{15}}, \bibinfo{pages}{2475--2486}
  (\bibinfo{year}{2016}).

\bibitem{219067:4921053}
\bibinfo{author}{Behera, B.~K.}, \bibinfo{author}{Banerjee, A.} \&
  \bibinfo{author}{Panigrahi, P.~K.}
\newblock \bibinfo{title}{Experimental realization of quantum cheque using a
  five-qubit quantum computer}.
\newblock \emph{\bibinfo{journal}{Quantum Inf. Process.}}
  \textbf{\bibinfo{volume}{16}}, \bibinfo{pages}{312} (\bibinfo{year}{2017}).

\bibitem{219067:4921054}
\bibinfo{author}{Mihara, T.}
\newblock \bibinfo{title}{Quantum identification schemes with entanglements}.
\newblock \emph{\bibinfo{journal}{Phys. Rev. A}}
  \textbf{\bibinfo{volume}{65}}, \bibinfo{pages}{052326}
  (\bibinfo{year}{2002}).

\bibitem{219067:4921055}
\bibinfo{author}{Hotta, M.} \& \bibinfo{author}{Ozawa, M.}
\newblock \bibinfo{title}{A protocol of quantum authentication with secure
  quantum passwords}.
\newblock In \emph{\bibinfo{booktitle}{AIP Conference Proceedings}}, vol.
  \bibinfo{volume}{1110}, \bibinfo{pages}{388--391} (\bibinfo{year}{2009}).

\bibitem{219067:4921056}
\bibinfo{author}{Garcia-Escartin, J.~C.} \& \bibinfo{author}{Chamorro-Posada,
  P.}
\newblock \bibinfo{title}{Simple quantum password checking}.
\newblock \emph{\bibinfo{journal}{Phys. Rev. A}}
  \textbf{\bibinfo{volume}{91}}, \bibinfo{pages}{062310}
  (\bibinfo{year}{2015}).

\bibitem{219067:4921057}
\bibinfo{author}{Su, S.-W.} \emph{et~al.}
\newblock \bibinfo{title}{Setting a disordered password on a photonic memory}.
\newblock \emph{\bibinfo{journal}{Phys. Rev. A}}
  \textbf{\bibinfo{volume}{95}}, \bibinfo{pages}{061805}
  (\bibinfo{year}{2017}).

\bibitem{219067:4921058}
\bibinfo{author}{Bennett, C.~H.}, \bibinfo{author}{DiVincenzo, D.~P.},
  \bibinfo{author}{Smolin, J.~A.} \& \bibinfo{author}{Wootters, W.~K.}
\newblock \bibinfo{title}{Mixed-state entanglement and quantum error
  correction}.
\newblock \emph{\bibinfo{journal}{Phys. Rev. A}}
  \textbf{\bibinfo{volume}{54}}, \bibinfo{pages}{3824} (\bibinfo{year}{1996}).

\bibitem{219067:4920990}
\bibinfo{author}{Berta, M.}, \bibinfo{author}{Wehner, S.} \&
  \bibinfo{author}{Wilde, M.~M.}
\newblock \bibinfo{title}{Entropic uncertainty and measurement reversibility}.
\newblock \emph{\bibinfo{journal}{New J. Phys.}}
  \textbf{\bibinfo{volume}{18}}, \bibinfo{pages}{073004}
  (\bibinfo{year}{2016}).

\bibitem{219067:4921059}
\bibinfo{author}{Alsina, D.} \& \bibinfo{author}{Latorre, J.~I.}
\newblock \bibinfo{title}{Experimental test of mermin inequalities on a
  five-qubit quantum computer}.
\newblock \emph{\bibinfo{journal}{Phys. Rev. A}}
  \textbf{\bibinfo{volume}{94}}, \bibinfo{pages}{012314}
  (\bibinfo{year}{2016}).

\bibitem{219067:4920991}
\bibinfo{author}{Wootton, J.~R.}
\newblock \bibinfo{title}{Demonstrating non-abelian braiding of surface code
  defects in a five qubit experiment}.
\newblock \emph{\bibinfo{journal}{Quantum Sci. Technol.}}
  \textbf{\bibinfo{volume}{2}}, \bibinfo{pages}{015006} (\bibinfo{year}{2017}).

\bibitem{219067:4920992}
\bibinfo{author}{Kalra, A.~R.}, \bibinfo{author}{Prakash, S.},
  \bibinfo{author}{Behera, B.~K.}, \bibinfo{author}{Panigrahi, P.}
  \emph{et~al.}
\newblock \bibinfo{title}{Experimental demonstration of the no hiding theorem
  using a 5 qubit quantum computer}.
\newblock \emph{\bibinfo{journal}{arXiv preprint arXiv:1707.09462}}
  (\bibinfo{year}{2017}).

\bibitem{219067:4920993}
\bibinfo{author}{Ghosh, D.}, \bibinfo{author}{Agarwal, P.},
  \bibinfo{author}{Pandey, P.}, \bibinfo{author}{Behera, B.~K.} \&
  \bibinfo{author}{Panigrahi, P.~K.}
\newblock \bibinfo{title}{Automated error correction in ibm quantum computer
  and explicit generalization}.
\newblock \emph{\bibinfo{journal}{arXiv preprint arXiv:1708.02297}}
  (\bibinfo{year}{2017}).

\bibitem{219067:4920994}
\bibinfo{author}{Gangopadhyay, S.}, \bibinfo{author}{Behera, B.~K.},
  \bibinfo{author}{Panigrahi, P.~K.} \emph{et~al.}
\newblock \bibinfo{title}{Generalization and partial demonstration of an
  entanglement based deutsch-jozsa like algorithm using a 5-qubit quantum
  computer}.
\newblock \emph{\bibinfo{journal}{arXiv preprint arXiv:1708.06375}}
  (\bibinfo{year}{2017}).

\bibitem{219067:4920995}
\bibinfo{author}{Wootton, J.~R.} \& \bibinfo{author}{Loss, D.}
\newblock \bibinfo{title}{A repetition code of 15 qubits}.
\newblock \emph{\bibinfo{journal}{arXiv preprint arXiv:1709.00990}}
  (\bibinfo{year}{2017}).

\bibitem{219067:4920996}
\bibinfo{author}{Huffman, E.} \& \bibinfo{author}{Mizel, A.}
\newblock \bibinfo{title}{Violation of noninvasive macrorealism by a
  superconducting qubit: Implementation of a leggett-garg test that addresses
  the clumsiness loophole}.
\newblock \emph{\bibinfo{journal}{Phys. Rev. A}}
  \textbf{\bibinfo{volume}{95}}, \bibinfo{pages}{032131}
  (\bibinfo{year}{2017}).

\bibitem{219067:4920997}
\bibinfo{author}{Sisodia, M.}, \bibinfo{author}{Shukla, A.} \&
  \bibinfo{author}{Pathak, A.}
\newblock \bibinfo{title}{Experimental realization of nondestructive
  discrimination of bell states using a five-qubit quantum computer}.
\newblock \emph{\bibinfo{journal}{Phys. Lett. A}}
  \textbf{\bibinfo{volume}{381}}, \bibinfo{pages}{3860--3874}
  (\bibinfo{year}{2017}).

\bibitem{219067:4920998}
\bibinfo{author}{Schuld, M.}, \bibinfo{author}{Fingerhuth, M.} \&
  \bibinfo{author}{Petruccione, F.}
\newblock \bibinfo{title}{Implementing a distance-based classifier with a quantum interference
  circuit}.
\newblock \emph{\bibinfo{journal}{EPL (Europhysics Letters)}}
  (\bibinfo{year}{2017}).

\bibitem{219067:4920999}
\bibinfo{author}{Majumder, A.}, \bibinfo{author}{Mohapatra, S.} \&
  \bibinfo{author}{Kumar, A.}
\newblock \bibinfo{title}{Experimental realization of secure multiparty quantum
  summation using five-qubit ibm quantum computer on cloud}.
\newblock \emph{\bibinfo{journal}{arXiv preprint arXiv:1707.07460}}
  (\bibinfo{year}{2017}).

\bibitem{219067:4921000}
\bibinfo{author}{Kandala, A.} \emph{et~al.}
\newblock \bibinfo{title}{Hardware-efficient variational quantum eigensolver
  for small molecules and quantum magnets}.
\newblock \emph{\bibinfo{journal}{Nature}} \textbf{\bibinfo{volume}{549}},
  \bibinfo{pages}{242} (\bibinfo{year}{2017}).

\bibitem{219067:4921001}
\bibinfo{author}{Li, R.}, \bibinfo{author}{Alvarez-Rodriguez, U.},
  \bibinfo{author}{Lamata, L.} \& \bibinfo{author}{Solano, E.}
\newblock \bibinfo{title}{Approximate quantum adders with genetic algorithms:
  An ibm quantum experience}.
\newblock \emph{\bibinfo{journal}{Quantum Measure. Quantum Metrol.}}
  \textbf{\bibinfo{volume}{4}}, \bibinfo{pages}{1--7} (\bibinfo{year}{2017}).

\bibitem{219067:4921002}
\bibinfo{author}{Sisodia, M.}, \bibinfo{author}{Shukla, A.},
  \bibinfo{author}{Thapliyal, K.} \& \bibinfo{author}{Pathak, A.}
\newblock \bibinfo{title}{Design and experimental realization of an optimal
  scheme for teleportation of an n-qubit quantum state}.
\newblock \emph{\bibinfo{journal}{Quantum Inf. Process.}}
  \textbf{\bibinfo{volume}{16}}, \bibinfo{pages}{292} (\bibinfo{year}{2017}).

\bibitem{219067:4921003}
\bibinfo{author}{Joy, D.}, \bibinfo{author}{Behera, B.~K.},
  \bibinfo{author}{Panigrahi, P.~K.} \emph{et~al.}
\newblock \bibinfo{title}{Experimental demonstration of non-local
  controlled-unitary quantum gates using a five-qubit quantum computer}.
\newblock \emph{\bibinfo{journal}{arXiv preprint arXiv:1709.05697}}
  (\bibinfo{year}{2017}).

\bibitem{219067:4929855}
\bibinfo{author}{Aharonov, Y.}, \bibinfo{author}{Albert, D.~Z.} \&
  \bibinfo{author}{Vaidman, L.}
\newblock \bibinfo{title}{How the result of a measurement of a component of the
  spin of a spin-1/2 particle can turn out to be 100}.
\newblock \emph{\bibinfo{journal}{Phys. Rev. Lett.}}
  \textbf{\bibinfo{volume}{60}}, \bibinfo{pages}{1351--1354}
  (\bibinfo{year}{1988}).

\end{thebibliography}
\end{document}